\documentclass[a4paper, 12 pt, twoside, reqno]{amsart}

\usepackage[english]{babel}
\usepackage[utf8]{inputenc}

\usepackage{setspace}
\usepackage[euler-digits,euler-hat-accent]{eulervm}

\usepackage{times}
\usepackage{amsmath}
\usepackage{amsfonts}
\usepackage{amssymb}
\usepackage{amsmath}
\usepackage{amsthm}
\usepackage{graphicx}
\usepackage{array}
\usepackage{color}
\usepackage{mathrsfs}
\usepackage{hyperref}
\usepackage{eucal}
\usepackage{esint} %%% for \fint
\usepackage{tikz}
\usepackage{upgreek}
\usepackage{enumitem}

\usepackage{tikz-cd}

\allowdisplaybreaks

\theoremstyle{plain}
\newtheorem{theorem}{Theorem}[section]
\newtheorem{corollary}[theorem]{Corollary}
\newtheorem{proposition}[theorem]{Proposition}
\newtheorem{lemma}[theorem]{Lemma}

\theoremstyle{definition}
\newtheorem{definition}[theorem]{Definition}

\theoremstyle{remark}
\newtheorem{remark}[theorem]{Remark}

\numberwithin{equation}{section}
\numberwithin{figure}{section}
\numberwithin{table}{section}

\newcommand{\R}{\mathbb{R}}
\newcommand{\N}{\mathbb{N}}
\newcommand{\C}{\mathbb{C}}                           

\newcommand{\Z}{\mathbb{Z}}

\newcommand{\s}[1]{\CMcal{#1}}
                  
\newcommand{\bb}[1]{\mathscr{#1}}
\newcommand{\rr}[1]{\mathfrak{#1}}
\newcommand{\n}[1]{\mathbb{#1}}

\newcommand{\expo}[1]{\,\mathrm{e}^{#1}\,}                 

\newcommand{\dd}{\,\mathrm{d}}
\newcommand{ \ii}{\,\mathrm{i}\,}

\newcommand{\virg}[1]{\lq\lq#1\rq\rq}                \newcommand{\ie}{\textsl{i.\,e.\,}}
\newcommand{\eg}{\textsl{e.\,g.\,}}

\newcommand{\etc}{\textsl{etc}.\,}

\newlength{\dhatheight}

\hypersetup{
pdftoolbar=true,        % show Acrobat’s toolbar?
pdfmenubar=true,        % show Acrobat’s menu?
pdffitwindow=true,     % window fit to page when opened
pdfstartview=true,    % fits the width of the page to the window
pdftitle={Topology-of-States},    % title
pdfauthor={Giuseppe De Nittis, Santiago G. Rendel},     % author
breaklinks=true, %permet le retour a la ligne dans les liens trop longs
colorlinks=true,       % false: boxed links; true: colored links
linkcolor=purple,         % color of internal links (change box color 
citecolor=teal, % color of links to bibliography
urlcolor=blue, %couleur des hyperliens
bookmarksopen=true, %si les signets Acrobat sont crees,
filecolor=magenta,      % color of file links
}

\begin{document}

\title[Topological phases of non-interacting systems]{
Topological phases of non-interacting systems:\\ A general approach based on  states}

\author[G. De~Nittis]{Giuseppe De Nittis}

\address[G. De~Nittis]{Facultad de Matemáticas \& Instituto de Física,
  Pontificia Universidad Católica de Chile,
  Santiago, Chile.}
\email{gidenittis@uc.cl}

%\author[S.~G. Rendel]{Santiago G. Rendel}
%\address[S.~G. Rendel]{Facultad de Matemáticas \& Instituto de Física,
%  Pontificia Universidad Católica de Chile,
%  Santiago, Chile.}
%\email{sgr@uc.cl}
%

\vspace{2mm}

\date{\today}

\begin{abstract}
In this work we provide a classification scheme for topological phases 
of certain systems whose observable algebra is described by a trivial $C^*$-bundles. 
The classification is based on the study of the homotopy classes of \emph{configurations}, which are maps from a \emph{quantum parameter space} to the space of pure states of a reference \emph{fiber}
$C^*$-algebra. Both the quantum parameter space and the fiber algebra
are naturally associated with the observable algebra.
A list of various examples described in the last section shows that  
the common classification scheme of non-interacting topological insulators of type A is recovered inside this new formalism.

 \medskip

\end{abstract}

\maketitle

\tableofcontents

%------%
\section{Introduction}\label{sec:Intr0}
The study and classification of topological states of matter is an interesting and attractive topic, that has garnered the attention of the mathematical physics community in recent years. The beginning of this story dates back to 1982, with the discovery of the topological nature of the quantum Hall effect (QHE) with the seminal works \cite{thouless-kohmoto-nightingale-nijs-82}. Systems showing  QHE were the first examples of a large class of systems today called \emph{topological insulators} (TI). The latter are materials that show topologically protected phases, and are classified in terms of certain fundamental symmetries (like time-reversal or particle hole). The first logical organization of these systems, called  was proposed by 
in \cite{altland-zirnbauer-97} (see also \cite{ryu-schnyder-furusaki-ludwig-10}), but it was only in 2009 with the breakthrough paper \cite{kitaev-09}, that Kitaev proposed a unified framework for the complete classification of all possible topological phases based on the use of \emph{$K$-theory}. This is the so called \emph{Kitaev's periodic table} of topological insulators. Today, the use of $K$-theory for the classification of TI is a commonly recognized and powerful tool, which in fact has its roots in the pioneering works by  Bellissard on the QHE \cite{bellissard-elst-schulz-baldes-94}.
 We will refer to \cite{prodan-schulz-baldes-book} for a general overview, and a rich bibliography,  on the extended and generalized application of 
 $K$-theory in the   context of  topological materials.

\medskip

The main limitation of the Kitaev's periodic table is that it only works for \emph{non-interacting} systems. This contrasts with recent advances in the proof of the QHE for  \emph{interacting} extended fermionic systems \cite{jaksic-ogata-pillet-06,hastings-michalakis-15,giuliani-mastropietro-porta-17,monaco-teufel-19,bachmann-bols-roeck-fraas-20}.
Therefore, the need arises to have a classification scheme of topological phases suitable for interacting systems. The main difficulty lies in the fact that $K$-theory is, ultimately, a classification scheme for gapped spectral projectors of the Hamiltonian describing the system. On the other hand, the  Hamiltonian, together with its spectral projections, 
 is not a prime concept, sometime not even well defined, in the context of extended interacting systems.
In this setting, the natural replacement for spectral projectors (or more in general density matrices) is the notion of states. This leads to the need of a classification scheme for topological phases of matter based on the states. 
Although recent progress in this direction has already been made \cite{Beaudry-hermele-moreno-etc-24,spiegel-pflaum-24,artymowicz-kapustin-24}, a complete and general picture is still missing. 
It is worth mentioning that there is a modern proposal to frame the classification problem of topological phases in an abstract , and somewhat universal, definition. According to the  \emph{Kitaev’s conjecture}  that gapped invertible phases of (not necessary) interacting systems form
a \emph{loop-spectrum} in the sense of homotopy theory. For more details and references on this idea the reader is referred  to \cite{Beaudry-hermele-moreno-etc-24}. Nevertheless, in our opinion,	there are fundamental questions that still deserve an answer, such as: \emph{What is the exact definition of equivalence of states? Which classes of states support a topological classification? Is it possible to redeem traces of $K$-theory in a classification scheme based on the notion of states?}

\medskip

In order to shed some light on these open problems, and without having the pretension  to solve the problem in general, we 
provide in this note a possible simplified classification scheme for topological phases of non-interacting systems based on the notion of state and aimed to recover track of the $K$-theory  
when applied to the well understood case of non-interacting type systems.

\medskip

 Let $X$ be a compact Hausdorff space playing the role of a \emph{quantum parameter} space. We will assume that our algebra of  observable $\bb{A}$ has the structure of a \emph{$C^*$-bundle}, meaning that there is a collection of $C^*$-algebras $\bb{A}_x$ parametrized by $x\in X$, and continuous surjective $\ast$-homomorphisms $\pi_x:\bb{A}\to \bb{A}_x$  such that the maps
$X\ni x\to\|\pi_x(\rr{a})\|\in\C$ are continuous for every $\rr{a}\in\bb{A}$. The maps $\pi_x$ are called \emph{localizing homomorphisms} and every $C^*$-algebra $\bb{A}_x$ describe the \emph{internal} observables of the \emph{elemental system} localized at $x$. In this sense $\bb{A}$ plays the role of the $C^*$-algebra of extended observable. It is worth emphasizing that the parameter $x$ can have the meaning of any relevant quantum number such as position, momentum, quasi-momentum \etc.
We will refer to  Section \ref{sec:C-bun} for more details about $C^*$-bundles. Let $\s{S}_{\bb{A}}$ be the state space of 
$\bb{A}$ and $\s{P}_{\bb{A}}\subset\s{S}_{\bb{A}}$ the subset of \emph{pure states} both endowed with the $\ast$-weak topology.
The following definition is inspired by certain ideas in \cite{Beaudry-hermele-moreno-etc-24}.
\begin{definition}[Configuration]
Let $\bb{A}$ be a   $C^*$-bundle over the compact Hausdorff space $X$. Let $C(X,\s{P}_{\bb{A}})$ be the space of continuous functions on $X$ with values on the pure states of  $\bb{A}$. Any $F\in C(X,\s{P}_{\bb{A}})$ will be called a \emph{configuration}  for $\bb{A}$.
\end{definition}

\medskip

Given the configuration $F\in C(X,\s{P}_{\bb{A}})$, and any positive normalized Borel measure $\mu\in\s{M}_{+,1}(X)$, one can define a state $\omega_{\mu,F}\in \s{S}_{\bb{A}}$ by the prescription
\begin{equation}\label{eq:-int-00}
\omega_{\mu,F}(\rr{a})\;:=\;\int_X\dd\mu(x)\; F(x)[\rr{a}]\;,\qquad \rr{a}\in \bb{A}
\end{equation}
where $F(x)\in \s{P}_{\bb{A}}$
denotes the evaluation of $F$ at $x$. The set
\[
{\rm Stat}(F)\;:=\;\{\omega_{\mu,F}\;|\; \mu\in\s{M}_{+,1}(X)\}\;\subset\;\s{S}_{\bb{A}}
\]
will be called the \emph{statistical ensamble} subordinate to $F$. This set
contains all  the statistical information of the system encoded in its configuration $F$. 

\medskip

For physical application it  makes sense to interpret each $F(x)$ as a property of the system localized at $x$. To do that let  $\s{P}_{\bb{A}_x}$ be the set of pure states of the $C^*$-algebra $\bb{A}_x$ localized at $x$. Let $\omega_x\in \s{P}_{\bb{A}_x}$  and consider the \emph{lift map} $\rr{i}(\omega_x):=\omega_x\circ\pi_x$. Then, it is relevant to observe that 
$\rr{i}:\s{P}_{\bb{A}_x}\to \s{P}_{\bb{A}}$ 
is an injection (Proposition \ref{pro inj}).

\begin{definition}[Localizable configuration]\label{def:LC}
An element  $F\in C(X,\s{P}_{\bb{A}})$ will be called a \emph{localizable configuration}  if $F(x)\in \rr{i}(\s{P}_{\bb{A}_x})$ for every $x\in X$. The set of localizable configurations will be denoted with $C_\s{L}(X,\s{P}_{\bb{A}})$.
\end{definition}

\medskip

According to the definition above if $F$ is a localizable configuration then for every $x\in X$ the evaluation $F(x)$ identifies a unique pure states of the localized system $\bb{A}_x$.
Therefore, a localizable configuration can be seen as a map which fixes to each $x$ a unique element of $\s{P}_{\bb{A}_x}$. This implies that \emph{globally} a localizable  configuration appears as a \virg{superposition} of distinct pure states
since states localized at distinct quantum numbers are indeed distinct. In this sense, the definition of localizable configuration 
can be interpreted as an incarnation of   the \emph{Pauli's exclusion principle} of quantum mechanics remodeled to fit in this abstract framework. 
Having this in mind it would be also appropriate to call \emph{fermionic} the configuration described in Definition \ref{def:LC}.

\medskip

Different configurations that can be deformed \virg{nicely} one into the other must belong to the same \emph{phase}. To make this idea rigorous let us introduce the following notion.
Two configurations $F,G\in C(X,\s{P}_{\bb{A}})$ are \emph{homotopy} equivalent if there exists a continuous (deformation) function $\phi:X\times[0,1]\to \s{P}_{\bb{A}}$ such that $\phi(x,0)=F(x)$ and $\phi(x,1)=G(x)$ for every $x\in X$. If this is the case we will use the notation $F\thicksim G$ to denote the homotopy equivalence between $F$ and $G$. The set of equivalence classes of  homotopy equivalent functions is usually  denoted with $[X,\s{P}_{\bb{A}}]$. 

\begin{definition}[Topological phases]\label{int:def_1}
A \emph{topological phase} of $\bb{A}$ is an equivalence class of configurations under the relation induce by the homotopy. The set of 
topological phases will be denoted with
\[
{\rm TF}(\bb{A})\;:=\;[X,\s{P}_{\bb{A}}]\;.
\]
We will refer ate the class $[F]\in {\rm TF}(\bb{A})$ as the
the topological phase represented by the  configuration $F$.
\end{definition}

\begin{remark}[Symmetries]
It make sense to generalize  Definition \ref{int:def_1} by replacing the ful set $\s{P}_{\bb{A}}$ by some subset $\s{P}_{\bb{A}}^{\n{G}}\subset \s{P}_{\bb{A}}$ defined by the invariance under some \emph{group of symmetry} $\n{G}$. However, in many situations
one has homeomorphisms of the type $\s{P}_{\bb{A}}^{\n{G}}\simeq\s{P}_{\bb{A}_{\n{G}}}$
where $\bb{A}_{\n{G}}$ is a new $C^*$-algebra determined by the symmetry group $\n{G}$. A situation of this type occurs for instance 
for the Weyl $C^*$-algebra where $\n{G}$ acts as the group of spatial translations \cite{denittis-gonzalez-25}. In this sense Definition \ref{int:def_1}
appears flexible enough to cover cases not immediately included in the scheme presented above. 
 \hfill $\blacktriangleleft$
\end{remark}

\medskip

It is important to notice that when the configurations $F$ and $G$ are in the same phase then, given any $\mu\in \s{M}_{+,1}(X)$, the states $\omega_{\mu,F}$ and $\omega_{\mu,G}$ can
be deformed into each other continuously with respect to $\ast$-weak topology (Proposition \ref{pr_sta_equiv}). This means that the statistics associated to $F$ and $G$ are equivalent under continuous deformations of the system. We will express this concept with the symbol ${\rm Stat}(F) \thicksim{\rm Stat}(G)$.

\medskip

One can be more restrictive about the definition of the topological phase by imposing the condition of localizability of the configurations. In this case two elements $F,G\in C_\s{L}(X,\s{P}_{\bb{A}})$ are equivalent in the class of   localizable configurations if there exists a continuous (deformation) function $\phi:X\times[0,1]\to \s{P}_{\bb{A}}$ such that $\phi(x,0)=F(x)$ and $\phi(x,1)=G(x)$ for every $x\in X$ and $\phi(\cdot, t)\in C_\s{L}(X,\s{P}_{\bb{A}})$ for every $t\in[0,1]$. 
The set of equivalence classes of  homotopy equivalent localizable configurations will be  denoted with $[X,\s{P}_{\bb{A}}]_\s{L}$. 

\begin{definition}[Localizable topological phases]
A \emph{localizable topological phase} of $\bb{A}$ is an equivalence class of localizable configurations under the relation induce by the homotopy. The set of localizable
topological phases will be denoted with
\[
{\rm LTF}(\bb{A})\;:=\;[X,\s{P}_{\bb{A}}]_\s{L}\;.
\]
\end{definition}

\medskip

In many physical applications it makes sense to consider  the 
 \emph{extended} system $\bb{A}$ as made of infinite copies of a given \emph{elemental system} whose \emph{internal} observables are represented by a fixed $C^*$-algebra $\bb{O}$. This translates into the condition  of considering   $C^*$-bundles  with a given \emph{typical fiber}
 $\bb{A}_x\simeq \bb{O}$.
 The simplest examples of this type of structure  are the trivial $C^*$-bundles
\begin{equation}\label{eq:-int-01}
\bb{A}\;:=\;C(X,\bb{O})\;\simeq\; C(X)\otimes\bb{O}\;.
\end{equation}
$C^*$-algebras of the type \eqref{eq:-int-01} are well studied object  and are also very common in physical applications. For more details we will refer to Section \ref{sec:C-bun}, and we will provide various concrete examples in Section \ref{sec:appl_syst}.

\medskip

The study of the space of pure states for a $C^*$-algebra of the type \eqref{eq:-int-01} is facilitate by the homeomorphism
$\s{P}_{\bb{A}}\simeq X\times \s{P}_{\bb{O}}$ (Proposition \ref{prop:prod_st}).
In view of that, localizable configurations correspond to sections of the trivial bundle $X\times \s{P}_{\bb{O}}\to X$, and in turn one gets the identification $C_\s{L}(X,\s{P}_{\bb{A}})\simeq C(X,\s{P}_{\bb{O}})$ 
(Proposition \ref{prop_idennt_OO}). These observations are preparatory for the following result.
\begin{theorem}\label{th:main1}
Let $\bb{A}$ be an extended system of the type \eqref{eq:-int-01}.
Then one has bijections
\[
{\rm LTF}(\bb{A})\;\simeq\;[X,\s{P}_{\bb{O}}]
\]
and 
\[
{\rm TF}(\bb{A})\;\simeq\;[X,X\times \s{P}_{\bb{O}}]\;=\;{\rm Deg}(X)\times{\rm LTF}(\bb{A})\;
\]
where ${\rm Deg}(X):=[X,X]$. 
\end{theorem}
\medskip

The details of the proof of the result above are postponed to  Section \ref{sec:th_proof}. Let us observe that the  set ${\rm Deg}(X)$ consists of the homotopy classes of the continuous functions $F:X\to X$. For that it makes sense to refer to ${\rm Deg}(X)$ as 
 the \emph{degree set} of $X$.
The class $d(F)\in {\rm Deg}(X)$ will be called the \emph{degree} of the configuration $F$.  

\medskip

In the special (but still very interesting) case where $\bb{O}=\bb{K}(\s{H})$ is the $C^*$-algebra of compact operators over some separable Hilbert space $\s{H}$,   one gets an homeomorphism $\s{P}_{\bb{K}}\simeq \n{P}(\s{H})_w$ where $\n{P}(\s{H})_w$ is the projective space of $\s{H}$ endowed with the weak topology of operators (Remark \ref{rk:ch_top} \& Lemma \ref{lem:comp}). Therefore, one ends with the following result which reproduce the topological classification of \emph{type A} non-interacting topological insulators \cite{altland-zirnbauer-97,kitaev-09,ryu-schnyder-furusaki-ludwig-10}.
\begin{theorem}\label{th:main2}
Let $X$ be a compact Hausdorff space  homotopy equivalent to a $CW$-complex.
Let $\bb{A}$ be an extended system of the type \eqref{eq:-int-01}
with typical fiber $\bb{K}(\s{H})$. Then
\[
{\rm LTF}(\bb{A})\;\simeq\;[X,\n{P}(\s{H})_w]\;\simeq\;{\rm Pic}(X)\;\stackrel{c_1}{\simeq}\;H^2(X,\Z)
\]
where ${\rm Pic}(X)$ is the \emph{Picard group} of  complex line bundles over $X$ and the map $c_1$ which provides the last bijection (indeed a group isomomorphism) is the \emph{first Chern class}.
\end{theorem}

\medskip

The result above has an immediate consequence when $X$ has the structure of a low dimensional $CW$-complex.
\begin{corollary}\label{coro:int}
Let $X$ be a finite $CW$-complex of dimension $d\leqslant 3$. Under the conditions of Theorem \ref{th:main2} one has that 
\[
{\rm LTF}(\bb{A})\;\simeq\;\widetilde{K}^0(X)
\]
where $\widetilde{K}^0(X)$  denotes the reduced topological complex $K$-group of $X$.
\end{corollary}

 \medskip
 
 This result shows how $K$-theory appears in the classification of 
  localizable topological phases of certain classes of physical systems. This connection will  be further explained in Section \ref{sec:appl_syst} with various specific examples which encompass  almost  all the  non-interacting models generally considered in condensed matter.

 \medskip  
  
 \noindent
{\bf Acknowledgements.}
GD's research is supported by the grant \emph{Fondecyt Regular - 1230032}.  GD would like to cordially thank Santiago G. Rendel  for many stimulating discussions about the subject of this work.

%------------------------------------------------------------------------%
\section{Algebras, states and topological phases}
\label{sec:alg-st}
%------%
\subsection{$C^*$-bundles}\label{sec:C-bun}
Let $X$ be a compact Hausdorff space and $C(X)$ the $C^*$-algebra of continuous functions over $X$. 
The preliminar concept we need is that of \emph{$C(X)$-algebra} initially introduced in \cite[Definition 1.5]{kasparov-88}. A $C(X)$-algebra is a $C^*$-algebra $\bb{A}$  endowed with a unital $\ast$-homomorphism  (called \emph{structure morphism}) 
$\theta:\bb{A}\to\bb{Z}(\bb{M}(\bb{A}))$ where $\bb{M}(\bb{A})$ is the 
\emph{multiplier algebra} of $\bb{A}$ and $\bb{Z}(\bb{M}(\bb{A}))$ denotes the \emph{center} of $\bb{M}(\bb{A})$.
Let us recall that the center $\bb{Z}(\bb{B})$ of any $C^*$-algebra $\bb{B}$ is the set of all the elements in  $\bb{B}$ commuting with the full algebra. The multiplier algebra
$\bb{M}(\bb{A})$ is the largest unital $C^*$-algebra that contains $\bb{A}$ as an \emph{essential} ideal. When $\bb{A}$ is itself unital then the definition above is simplified by the equality  $\bb{M}(\bb{A})=\bb{A}$.

\medskip

For any $x\in X$ let $C_0(X/\{x\})$ be the  ideal of continuous
functions vanishing on $\{x\}$. Then $\bb{I}_x:=\theta(C_0(X/\{x\})) \bb{A}$
is a closed two-sided ideal of $\bb{A}$ by \emph{Cohen’s
factorization theorem} \cite[Theorem  II.5.3.7]{blackadar-06}.
The quotient $\bb{A}_x:=\bb{A}/\bb{I}_x$ is a $C^*$-algebra called the \emph{fiber} of $\bb{A}$ over the point $x$, and the associated \emph{quotient map} (a $\ast$-homomorphism) will be denoted with $\pi_x:\bb{A}\to\bb{A}_x$. In view of this fiber decomposition one may think
to each $\rr{a}\in\bb{A}$ as a function $\rr{a}:X\to\prod_{x\in X}\bb{A}_x$ such that $\rr{a}(x):=\pi_x(\rr{a})$. A general result is that the map $x\mapsto\|\rr{a}(x)\|$ is upper semicontinuous 
for all $\rr{a}\in\bb{A}$ \cite[Lemma 2.1]{dadarlat-09} .

\medskip

A $C(X)$-algebra such that the map $x\mapsto\|\rr{a}(x)\|$ is continuous for all $\rr{a}\in\bb{A}$ is called 
a \emph{$C^*$-bundle over $X$} \cite[Definition 2.2]{blanchard-kirchberg-04}. Interestingly, a $C^*$-algebra $\bb{A}$ is a $C^*$-bundle if and only if it coincides with the $C^*$-algebra of continuous sections of a continuous field of $C^*$-algebras over $X$ in
the sense of Dixmier \cite[Definition 10.3.1]{dixmier-77}.
For this equivalence the reader is referred to  \cite{nilsen-88,blanchard-kirchberg-04}.

\medskip

Let $\bb{O}$ be a  $C^*$-algebra. A  $C^*$-bundle $\bb{A}$ has \emph{typical fiber} $\bb{O}$ if there are isomorphisms $\bb{A}_x\simeq \bb{O}$ for every $x\in X$. Let
 $C(X,\bb{O})$ be the $C^*$-algebra of continuous function form $X$ to $\bb{O}$. It is straightforward check to see that $C(X,\bb{O})$ is an example of $C^*$-bundle over $X$ with typical fiber $\bb{O}$. A $C^*$-bundle $\bb{A}$ is \emph{trivial} (with typical fiber $\bb{O}$) if there exists a $\ast$-isomorphism of $C^*$-algebras $\bb{A}\simeq C(X,\bb{O})$. In the literature there exists a  similar notion of local triviality but this has no relevance for the purposes of this work.

\medskip

A trivial $C^*$-bundle $\bb{A}$ can be usefully represented as a tensor product. In fact  it holds true that
\[
\bb{A}\;\simeq\;C(X,\bb{O})\;\simeq\; C(X)\otimes\bb{O}\;.
\]
The second isomorphism is a classical result in the theory of tensor products of $C^\ast$-algebras and there is
no ambiguity in the specification of the type of tensor product since $C(X)$ is an abelian, hence nuclear, $C^*$-algebra  \cite[Proposition T.5.21 \& Corollary T.6.17]{wegge-olsen-93}.

\medskip

Let $\bb{K}(\s{H})\subset\bb{B}(\s{H})$ the the $C^*$-algebra of compact operators over the separable Hilbert space $\s{H}$. Trivial $C^*$-bundles with fiber 
$\bb{K}(\s{H})$, \ie 
\[
\bb{A}\;\simeq\;C(X,\bb{K}(\s{H}))\;\simeq\; C(X)\otimes \bb{K}(\s{H})\;,
\]
are special examples of \emph{continuous-trace $C^*$-algebras} 
\cite[Section 5.2]{raeburn-williams-98}.
For this type of algebras there is a homeomorphism $X\simeq{\rm Spec}(\bb{A})$ where ${\rm Spec}(\bb{A})$ denotes the \emph{spectrum} of  
$\bb{A}$, \ie the set of equivalence classes of irreducible representations of $\bb{A}$ with the induced hull-kernel topology \cite[Definition A.21 \& Example A.24]{raeburn-williams-98}.

%------%
\subsection{States  and pure states}
For the benefit of the reader we will sketch some of the most relevant concepts a bout  states of a $C^*$-algebra. For a more complete
presentation we will refer to \cite[Section 2.3.2]{bratteli-robinson-87}.

\medskip

Let $\bb{A}$ be a   $C^*$-algebra. A \emph{state} $\omega$ on $\bb{A}$ is a positive linear functional with norm $\|\omega\|=1$ .
The state space of $\bb{A}$ will be denotes with
 $\s{S}_\bb{A}$. A state is called \emph{pure} if it cannot be decomposed in the convex combination of two non-zero distinct states.
 The symbol  $\s{P}_\bb{A}$ will be used for
the subset of pure states of $\bb{A}$. The space $\s{S}_\bb{A}$ can be topologized with different topologies. For many reasons it is useful to use  the $\ast$-weak topology, that is the topology  of point-wise convergence of nets on  $\s{S}_\bb{A}$.
A basis of this topology is provided by the family of neighborhoods of any $\omega\in \s{S}_\bb{A}$ defined by
\[
\mathtt{U}_{\{\rr{a}_1,\ldots,\rr{a}_n\},\varepsilon}(\omega)\;:=\;\left\{\omega'\in\s{S}_\bb{A}\;|\; |\omega(\rr{a}_i)-\omega'(\rr{a}_1)| < \varepsilon\;, \quad i=1,\ldots,n\right\}
\]
and indexed by $\varepsilon>0$ and finite sets of elements $\{\rr{a}_1,\ldots,\rr{a}_n\}\subset\bb{A}$. When $\bb{A}$ is unital then $\s{S}_\bb{A}$ is a compact Hausdorff space with respect to the $\ast$-weak topology. Moreover, it is a convex space whose set of extremal points coincides with $\s{P}_\bb{A}$, and the full space  $\s{S}_\bb{A}$ can be obtained as the closure of  the convex envelope of $\s{P}_\bb{A}$ \cite[Theorem 2.3.15]{bratteli-robinson-87}. When $\bb{A}$ does not contain the identity then  $\s{S}_\bb{A}$ cannot be compact. Moreover, it can also fail to be  locally compact  in general.
Finally,
the $\ast$-weak topology on $\s{S}_\bb{A}$ is metrizable when $\bb{A}$ is separable \cite[Section 4.1.4]{bratteli-robinson-87}. 

\medskip

Given any subset $\s{K}\subseteq\s{S}_\bb{A}$ we will see  $\s{K}$ as a topological space endowed with the subspace (or induced) topology. It turn out that  $\s{K}$ is automatically Hausdorff. When $\s{K}$ is closed inside $\s{S}_\bb{A}$ and there is a unit, then $\s{K}$ is again a compact space with respect to the induced topology. However, this is not the usual case. It is known that $\s{P}_\bb{A}$ is a Baire space (with respect to the  the subspace topology) \cite[Corollary 5.2.8]{saito-wright-15} but in general it may fail to be closed in $\s{S}_\bb{A}$. 

\medskip

When $\bb{C}$ is a commutative unital $C^*$-algebra $X=\s{P}_\bb{C}$ is a compact Hausdorff space and there is an isomorphism $\bb{C}\simeq C(X)$ called the \emph{Gelfand transform} \cite[Theorem 2.1.11A]{bratteli-robinson-87}. In particular this provides a bijection $\delta:X\to \s{P}_{C(X)}$ which identified each $x\in X$ with the \emph{evaluation} $\delta_x$ defined by $\delta_x(f):=f(x)$ for every $f\in C(X)$. The following result is often sated in the literature (see for instance \cite[Exercise 1.2]{gracia-varilly-figueroa-01} or \cite[Example A. 16]{raeburn-williams-98}). Let us just sketch the proof.
\begin{lemma}\label{lem:homCX}
Let $X$ be a compact Hausdorff space. The map  $\delta:X\to \s{P}_{C(X)}$ is a homeomorphism.
\end{lemma}
\proof
Let us start by the known fact that $\delta$ is a bijection. To prove that it is continuous let $\{x_\alpha\}\subset X$ by any net converging to $x_0$. Then, for any  $f\in C(X)$ one has that $f(x_\alpha)\to f(x_0)$ in view of the continuity. However, this means that $\delta_{x_\alpha}(f) \to\delta_{x_0}(f)$ for every $f\in C(X)$. The latter is the same that the net $\{\delta_{x_\alpha}\}\subset  \s{P}_{C(X)}$ converges to $\delta_{x_0}$ in the $\ast$-weak topology. 
Thus, $\delta$ is a continuous bijection. Since both spaces are compact and Hausdorff, then $\delta$ is a homeomorphism.
\qed

\medskip

As a second case of interest let us describe the space of pure states 
of the $C^*$-algebra $\bb{K}(\s{H})$ of compact operators over the separable Hilbert space $\s{H}$. For that let
\begin{equation}\label{eq:gra_1k}
\n{P}(\s{H})\;:=\;\left\{\rr{p}\in\bb{B}(\s{H})\;|\; \rr{p}=\rr{p}^*=\rr{p}^2\;,\;\;{\rm Tr}_{\s{H}}(P)=1 \right\}
\end{equation}
be the space of rank $1$ ortogonal projections of $\s{H}$. 
This space is know as \emph{projective space} of $\s{H}$.
Consider the map $\tau:\n{P}(\s{H})\to \s{P}_\bb{K}$ defined by $\tau_\rr{p}(\rr{a}):={\rm Tr}_{\s{H}}(\rr{pa})$ for every $\rr{a}\in \bb{K}(\s{H})$.
The fact that $\tau$ is a bijection is a well known fact in the literature \cite[Example 5.1.1]{murphy-90}. However, we need a more precise description. 
For that we need to   equip $\n{P}(\s{H})$ with the weak topology of operator induced by the inclusion in the algebra of bounded operators $\bb{B}(\s{H})$. We will use the notation $\n{P}(\s{H})_w$
to emphasize the choice of the topology.
The  topological space $\n{P}(\s{H})_w$  has been studied in \cite{shubin-96}.
\begin{remark}[The choice of the topology]\label{rk:ch_top}
The choice of the weak topology for $\n{P}(\s{H})$ might seem unnatural at first sight. Although the uniform topology might seem more natural, it is easy to be convinced that it is not the right topology to represent the space of pure states. Let $\n{P}(\s{H})_u$ be the projective space endowed with the uniform topology inherited from $\bb{B}(\s{H})$. According to \cite[Lemma 1.1]{shubin-96}
$\n{P}(\s{H})_u$ is a closed space. On the other hand $\s{P}_\bb{K}$ is not closed with respect to the $\ast$-weak topology. To see this let $\{e_n\}_{n\in\N}$ be an orthogonal basis of $\s{H}$ and $\rr{p}_n$ the projection that projects along $e_n$. Consider the family of pure states $\{\omega_n\}_{n\in\N}\subset \s{P}_\bb{K}$ defined by $\omega_n(\rr{a}):={\rm Tr}_{\s{H}}(\rr{p}_n\rr{a})$. It is not hard to see that $\lim_{n\to\infty}\omega_n(\rr{a})=0$ for every $\rr{a}\in \bb{K}(\s{H})$. In fact this is evidently true for finite rank operators, and then for density for every compact operator. As a consequence  $\omega_n\to 0$ in the $\ast$-weak topology which shows that $\s{P}_\bb{K}$  cannot be closed.
 \hfill $\blacktriangleleft$
\end{remark}

\begin{lemma}\label{lem:comp}
Let $\s{H}$ be a separable Hilbert space.
The map  $\tau:\n{P}(\s{H})_w\to \s{P}_\bb{K}$ is a homeomorphism.
\end{lemma}
\proof
The fact that $\tau$ is a bijection is well known, see   for instance  \cite[Example 5.1.1]{murphy-90} or \cite[Section 10.2]{barata-brum-chabu-21}. 
Since $\n{P}(\s{H})_w$ is metrizable \cite[Corollary 2.5]{shubin-96}, one
can check continuity on sequences. We will prove initially a stronger result. If  $\{\rr{p}_n\}_{n\in\N}\subset \n{P}(\s{H})_w$ is a sequence that converges (weakly) to $\rr{p}_*\in \n{P}(\s{H})_w$ then
\begin{equation}\label{eq:red}
\lim_{n\to\infty}{\rm Tr}_{\s{H}}(\rr{p}_n \rr{a})\;=\;{\rm Tr}_{\s{H}}(\rr{p}_*\rr{a})
\end{equation}
for every $\rr{a}\in \bb{B}(\s{H})$. Before proving \eqref{eq:red}, let us observe that in view of the inclusion  $\bb{K}(\s{H})\subset \bb{B}(\s{H})$, it implies that
\[
\lim_{n\to\infty}\tau_{\rr{p}_n}(\rr{a})\;=\;\tau_{\rr{p}_\ast}(\rr{a})
\]
for every $\rr{a}\in\bb{K}(\s{H})$. This 
shows that $\tau_{\rr{p}_n}\to \tau_{\rr{p}_\ast}$ in the $\ast$-weak topology, and in turn  $\tau$ is a continuous bijection.
To prove \eqref{eq:red}
let us start with $\rr{u}\in\bb{B}(\s{H})$ any unitary operator. Observe that 
\[
\big|{\rm Tr}_{\s{H}}(\rr{p}_n\rr{u})-{\rm Tr}_{\s{H}}(\rr{p}_*\rr{u})\big|\;\leqslant\;\|\rr{p}_n\rr{u}-\rr{p}_*\rr{u}\|_1\;=\;\|\rr{q}_n-\rr{q}_\ast\|_1
\]
where $\|\cdot\|_1$ denotes the norm of the Schatten ideal $\bb{L}^1(\s{H})$ of trace class operators and $\rr{q}_n:=\rr{p}_n\rr{u}$ for every $n\in \N\cup\{\ast\}$. One has that $\rr{q}_n\to \rr{q}_{*}$  in the weak topology.
Moreover, $|\rr{q}_n|=\rr{u}^*\rr{p}_n\rr{u}$ and $|\rr{q}_n^*|=\rr{p}_n$ for every $n\in \N\cup\{\ast\}$,
showing that also $|\rr{q}_n|\to|\rr{q}_{*}|$ and $|\rr{q}_n^*|\to|\rr{q}_{*}^*|$ in the weak topology. Finally $\|\rr{q}_n\|_1={\rm Tr}_{\s{H}}(\rr{u}^*\rr{p}_n\rr{u})=1$
 for every $n\in \N\cup\{\ast\}$. Since all the conditions of \cite[Theorem 2.20]{simon-05} are verified, one gets that $\|\rr{q}_n-\rr{q}_{*}\|_1\to 0$ when $n\to \infty$, and in turn 
\[
\lim_{n\to\infty}{\rm Tr}_{\s{H}}|(\rr{p}_n\rr{u})-{\rm Tr}_{\s{H}}(\rr{p}_*\rr{u})|\;=\;0
\]
for every unitary $\rr{u}$. Since every $\rr{a}\in\bb{B}(\s{H})$ 
can be written as the sum of  at most four unitaries $\rr{a}=\sum_{j=1}^4 a_j\rr{u}_j$ with $a_j\in\C$ bounded by $|a_j|\leqslant \|\rr{a}\|/2$
\cite[Lemma 2.2.14]{bratteli-robinson-87} one obtains by linearity \eqref{eq:red}.
The last task is to prove that the inverse $\tau^{-1}:\s{P}_\bb{K}\to \n{P}(\s{H})_w$  is also continuos. Since $\bb{K}(\s{H})$ is separable the space $\s{P}_\bb{K}$ is metrizable and one can check again continuity on sequences.
Therefore let $\{\omega_n\}_{n\in\N}\subset \s{P}_\bb{K}$
be a sequence converging to  $\omega_\ast\in \s{P}_\bb{K}$ in the $\ast$-weak topology. Let $\rr{p}_n:=\tau^{-1}(\omega_n)$ and $\rr{p}_\ast:=\tau^{-1}(\omega_\ast)$. Let $\rr{e}_{f,g}\in \bb{K}(\s{H})$ be the rank one operator defined by $\rr{e}_{f,g}(h):=\langle g,h\rangle f$ for every $f,g,h\in\s{H}$. Then
\[
\begin{aligned}
0\;&=\;\lim_{n\to\infty}[\omega_n(\rr{e}_{f,g})-\omega_\ast(\rr{e}_{f,g})]\\
&=\;\lim_{n\to\infty}{\rm Tr}_{\s{H}}[(\rr{p}_n-\rr{p}_\ast)\rr{e}_{f,g}]\;=\;\langle g,(\rr{p}_n-\rr{p}_\ast)f\rangle
\end{aligned}
\]
for every $f,g\in\s{H}$. Therefore,  $\tau^{-1}(\omega_n)\to \tau^{-1}(\omega_\ast)$ in the weak topology, proving the continuity of $\tau^{-1}$.\qed

\begin{remark}[Regular vs. irregular states]\label{rk:reg_st}
Let $\bb{B}(\s{H})$ be the full $C^*$-algebra  of bounded linear operators on the separable Hilbert space $\s{H}$. If $\rr{p}\in \n{P}(\s{H})_w$ then the map  $\tau_\rr{p}(\rr{a}):={\rm Tr}_{\s{H}}(\rr{pa})$ is  well defined for every $\rr{a}\in \bb{B}(\s{H})$ and provides a state  $\bb{B}(\s{H})$. It is indeed a pure state
as proved in \cite[Theorem 2.8]{araki-99} or \cite[Theorem 3]{barata-brum-chabu-21}. A look at the proof of Lemma \ref{lem:comp} shows that the map $\tau:\n{P}(\s{H})_w\to \s{P}_\bb{B}$ is still a continuous injection. However, it fails to be surjective when $\s{H}$ is infinite dimensional.
 Indeed, it is easy to provide examples of pure states $\omega \in\s{P}_\bb{B}$ such that $\omega(\rr{b})=0$ for all $\rr{b}\in \bb{K}(\s{H})$, and in turn that are not induced by a one-dimensional projection \cite[Section 10.1]{barata-brum-chabu-21}. The subset of pure states of $\bb{B}(\s{H})$ induced by projection is characterized 
by the    \emph{normality} condition as proved in  \cite[Theorem 3]{barata-brum-chabu-21} or \cite[Theorem 10]{barata-brum-chabu-21}. Let $\s{P}_{\bb{B},0}\subset \s{P}_\bb{B}$ be the subset of pure and \emph{normal} states. We will refer for short to the elements in $\s{P}_{\bb{B},0}$ as the \emph{regular states}. With a minimal amount of  modification the argument of the proof of 
Lemma \ref{lem:comp} implies that $\tau:\n{P}(\s{H})_w\to \s{P}_{\bb{B},0}$ is a homeomorphism. The set of pure states in $\s{P}_\bb{B}\setminus \s{P}_{\bb{B},0}$, which is characterized by the vanishing  on compacts, will be called the space of \emph{irregular} states.
 \hfill $\blacktriangleleft$
\end{remark}

\begin{remark}[Unitalization of the compacts]\label{rk:unit_comp}
Let $\bb{K}(\s{H})^+:=\bb{K}(\s{H})+\C {\bf 1}$ be the minimal unitalization of the $C^*$-algebra of compact operators
\cite[Section 2.3]{raeburn-williams-98}. Naturally, all the states $\tau_\rr{p}(\rr{a}):={\rm Tr}_{\s{H}}(\rr{pa})$ are still pure states for $\bb{K}(\s{H})^+$ (see \eg \cite[Theorem 4]{barata-brum-chabu-21})
However, there is a one more \emph{non regular} (non normal) state
defined by $\tau_0(\rr{a}+ z{\bf 1})=z$ for every $\rr{a}\in \bb{K}(\s{H})$ and $z\in\C$.
It turns out that $\s{P}_{\bb{K}^+}=\s{P}_\bb{K}\cup\{\tau_0\}$
and this is indeed a compactification of $\s{P}_\bb{K}$. With a minimal modification of the proof of Lemma \ref{lem:comp} one can prove the homeomorphism  ${\n{P}(\s{H})_w^+} \simeq \s{P}_{\bb{K}^+}$
where  ${\n{P}(\s{H})_w^+}:={\n{P}(\s{H})_w}\cup\{0\}$. In particular one has the interesting equality
${\n{P}(\s{H})_w^+} =\overline{\n{P}(\s{H})_w}\cap {\rm Proj}(\s{H})$ where in the right-hand side one has  the intersection of the weak closure of ${\n{P}(\s{H})_w}$
inside the bounded operators with the set of projections.
\cite[Proposition 3.4 (ii)]{shubin-96}.
 \hfill $\blacktriangleleft$
\end{remark}

%------%
\subsection{Statistical ensamble}
Given   $F\in C(X,\s{P}_{\bb{A}})$ and  $\mu\in\s{M}_{+,1}(X)$, let  $\omega_{\mu,F}$ be  the quantity  defined by \eqref{eq:-int-00}.
The continuity of the maps $F$ means that $X\ni x\mapsto F(x)[\rr{a}]\in\C$ is a continuous maps for every $\rr{a}\in\bb{A}$. Therefore, the integral which defines  $\omega_{\mu,F}$ makes sense for every $\rr{a}\in\bb{A}$. Linearity, positivity and normalization are easily verifications. Therefore, one gets that $\omega_{\mu,F}\in \s{S}_{\bb{A}}$  for every  $\mu\in\s{M}_{+,1}(X)$, and the definition of 
${\rm Stat}(F)$ is well posed. For the next result we will need one more concept. Let $\omega_{0},\omega_{1}\in \s{S}_{\bb{A}}$ be two states. We will say that $\omega_{0}$ and $\omega_{1}$ are \emph{path-equivalent} if there exists   a continuous map
$[0,1]\ni t\mapsto \omega_t\in \s{S}_{\bb{A}}$ (with respect to the $\ast$-weak topology) joining $\omega_{0}$ with $\omega_{1}$.
\begin{proposition}\label{pr_sta_equiv}
Let $F_0 \thicksim F_1$ be two homotopy equivalent configurations in $C(X,\s{P}_{\bb{A}})$ and   $\mu\in\s{M}_{+,1}(X)$ any measure. Then, the associated states  $\omega_{\mu,F_0}$ and $\omega_{\mu,F_1}$ are {path-equivalent}.
\end{proposition}
\proof
Let $\phi$ be the homotopy between $F_0$ and $F_1$, and for every $t\in[0,1]$ let $\omega_{\mu,t}\in \s{S}_{\bb{A}}$ be the element defined by
\[
\omega_{\mu,t}(\rr{a})\;:=\;\int_X\dd\mu(x)\; \phi(x,t)[\rr{a}]
\]
for every $\rr{a}\in \bb{A}$. By definition $\omega_{\mu,0}=\omega_{\mu,F_0}$ and 
$\omega_{\mu,1}=\omega_{\mu,F_1}$. Moreover, let $t_0\in[0,1]$ any number 
and observe that 
\begin{align*}
    \left|\omega_{\mu,t}(\rr{a}) - \omega_{\mu,t_0}(\rr{a}) \right| \;
    &\leqslant\; \int_{X} \dd\mu(x) \big| \phi(x,t)[\rr{a}] - \phi(x,t_0)[\rr{a}]\big|\;.
\end{align*}
Since $\phi(x,t)[\rr{a}] \to \phi(x,t_0)[\rr{a}]$ when $t\to t_0$ for all $\rr{a}\in \bb{A}$ in view of the 
continuity of $\phi$, an application of the dominated convergence theorem provides
\[
\lim_{t\to t_0} \left|\omega_{\mu,t}(\rr{a}) - \omega_{\mu,t_0}(\rr{a}) \right|\;=\;0\;.
\]
Therefore, the map $t\mapsto \omega_{\mu,t}$ is continuous at each point $t_0$ in the $\ast$-weak topology. This is the path which joins $\omega_{\mu,F_0}$ with $\omega_{\mu,F_1}$. 
\qed

\medskip

According two the previous result if  $F_0 \thicksim F_1$ then any element of ${\rm Stat}(F_0)$ is path-connected to the respective element of  
${\rm Stat}(F_1)$ for each choice of the measure $\mu$. 
When this happens we will say that the two statistical ensambles are equivalent, in symbol ${\rm Stat}(F_0)\thicksim{\rm Stat}(F_1)$.

\begin{remark}[From statistic to homotopy]\label{rk:stat-hom}
For the opposite of Proposition \ref{pr_sta_equiv} to be true  more conditions are necessary. Let $F_0, F_1\in C(X,\s{P}_{\bb{A}})$ be such that ${\rm Stat}(F_0)\thicksim{\rm Stat}(F_1)$.
We want to find conditions for the existence of a homotopy between $F_0$ and $F_1$. For every $x\in X$
let $\delta_x\in\s{M}_{+,1}(X)$ be te related \emph{delta measure}. Then,
by hypothesis $\omega_{\delta_x,F_0}$ and $\omega_{\delta_x,F_1}$ 
are connected by a path of states $[0,1]\ni t \mapsto \phi(x,t)\in \s{S}_{\bb{A}}$. Observe that by definition $\omega_{\delta_x,F_0}(\rr{a})=F(x)[\rr{a}]$ for every $\rr{a}\in \bb{A}$. This means that 
$\phi(x,0)=\omega_{\delta_x,F_0}=F_0(x)$ for every $x\in X$. Similarly,
one obtains that $\phi(x,1)=\omega_{\delta_x,F_1}=F_1(x)$. 
This shows that the map $\phi$ is the natural candidate for the homotopy.  
However, for this to be the case, two extra conditions are necessary: (i) the map $\phi$ must take values in the pure states $\s{P}_{\bb{A}}\subset \s{S}_{\bb{A}}$; (ii) the map $\phi:X\times [0,1]\to \s{P}_{\bb{A}}$ must be jointly continuous, while the hypotheses only guarantee the continuity of the restrictions $t\mapsto\phi(x,t)$ for each fixed $x\in X$.
 \hfill $\blacktriangleleft$
\end{remark}

%------%
\subsection{Pure states of a trivial $C^*$-bundle}
In this section we will describe de space of pure states of a 
trivial $C^*$-bundle. We will use the isomorphism $\bb{A} \simeq C(X)\otimes\bb{O}$ to exploit the tensor product structure.
For a detailed treatment of the theory tensor products of $C^*$-algebras we will refer to \cite[Chapter 6]{murphy-90}, \cite[Appendix T]{wegge-olsen-93} or  \cite[Appendix B]{raeburn-williams-98}.

\medskip

Let us recall first of all that $C(X)$ is a \emph{nuclear} $C^*$-algebra since it is  abelian \cite[Theorem 6.4.15]{murphy-90},
and by definition of nuclearity \cite[p. 193]{murphy-90} there is a unique norm which defines the tensor product $C(X)\otimes\bb{O}$. For instance one can think in  the \emph{spatial norm} for convenience (a characterization is given in \cite[Theorem 6.4.2]{murphy-90}).   Given states $\mu\in \s{S}_{C(X)}$ and $\omega\in \s{S}_{\bb{O}}$ one can define the product functional $\mu\otimes\omega:C(X)\otimes\bb{O}\to\C$ initially defined on simple tensors by $\mu\otimes\omega(f\odot \rr{a}):=\mu(f)\omega(\rr{a})$
for every $f\in C(X)$ and $\rr{a}\in \bb{O}$. Then $\mu\otimes\omega$ extends by linearity and density  to a unique state of $C(X)\otimes\bb{O}\to\C$ \cite[Corollary 6.4.3]{murphy-90}. Therefore, there  is a well defined mapping 
\[
\wp\;:\;\s{S}_{C(X)}\times \s{S}_{\bb{O}}\;\longrightarrow\; \s{S}_{C(X)\otimes\bb{O}}
\]
given by $\wp(\mu,\omega):=\mu\otimes\omega$. The injectivity of   
$\wp$ can be tested using the identity (or an approximate identity). In fact   $\wp(\mu,\omega)=\wp(\mu',\omega')$ implies that $\mu(f)\omega(\rr{a})=\mu'(f)\omega'(\rr{a})$ for every $f\in C(X)$ and $\rr{a}\in \bb{O}$. Replacing $f$ with the identity $1(x)=1$ 
and using the normalization of $\mu$ and $\mu'$ one gets $\omega(\rr{a})=\omega'(\rr{a})$ for every $\rr{a}\in \bb{O}$, and in turn $\omega=\omega'$. The equality $\mu=\mu'$ follows from an adaption of the same strategy. We are interested in the restriction of the map $\wp$ to pure states. 
\begin{lemma}
The map  $\wp$ defined above induces a bijection
\begin{equation}\label{eq:theta_map}
\wp\;:\;\s{P}_{C(X)}\times \s{P}_{\bb{O}}\;\longrightarrow\; \s{P}_{C(X)\otimes\bb{O}}\;.
\end{equation}
\end{lemma}
\proof
The fact  $\mu\in \s{P}_{C(X)}$ and $\omega\in \s{P}_{\bb{O}}$ 
implies that $\mu\otimes\omega\in \s{P}_{C(X)\otimes\bb{O}}$ is proved in \cite[Theorem 6.4.13]{murphy-90} (a simplified argument specifically valid for the spatial norm is contained in the proof of \cite[Theorem B.37]{raeburn-williams-98}). The map \eqref{eq:theta_map} is also surjective. First of all if $\eta\in\s{S}_{C(X)\otimes\bb{O}}$ one can define two elements $\mu_\eta\in \s{S}_{C(X)}$ and $\omega_\eta\in \s{S}_{\bb{O}}$ according to the prescriptions $\mu_\eta(f):=\lim_{n\to\infty}\eta(f\odot \rr{e}_n)$ and 
$\omega_\eta(\rr{a}):=\eta(1\odot \rr{a})$
where $\rr{e}_n$ is an approximated identity for $\bb{O}$ and and taking advantage of the fact that $C(X)$ is unital. 
If  $\eta\in\s{P}_{C(X)\otimes\bb{O}}$ is a pure state then 
$\mu_\eta\in \s{P}_{C(X)}$ is a pure state in view of the commutativity of ${C(X)}$ (as showed in the first part of the proof of \cite[Theorem 6.4.13]{murphy-90}). Therefore, this is sufficient to shows that also   $\omega_\eta\in \s{P}_{\bb{O}}$ is a pure state  and  $\eta= \mu_\eta\otimes \omega_\eta=\wp(\mu_\eta,\omega_\eta)$
\cite[Lemma 6.4.14]{murphy-90}. 
\qed

\medskip

The next step consists in the study of the topological property of the map \eqref{eq:theta_map}.

\begin{proposition}\label{prop:prod_st}
Let $\bb{A}\simeq C(X)\otimes\bb{O}$. Then there is a homeomorphism
\[
\s{P}_{\bb{A}}\;\simeq\; X\times \s{P}_{\bb{O}}\;.
\]
\end{proposition}
\proof
The main step of the proof is to check that the bijective map $\wp$ in \eqref{eq:theta_map} provides a homeomorphism when the spaces are endowed with  the $\ast$-weak topology.
Let $\{\mu_\alpha\}\subset \s{P}_{C(X)}$ be a net converging to $\mu_\ast\in \s{P}_{C(X)}$, and $\{\omega_\beta\}\subset \s{P}_{\bb{O}}$ be a net converging to $\omega_\ast\in \s{P}_{\bb{O}}$. Then $\wp(\mu_\alpha,\omega_\beta)(f\odot \rr{a})=\mu_\alpha(f)\omega_\beta(\rr{a})$ converges to $\mu_\ast(f)\omega_\ast(\rr{a})$ for every $f\in C(X)$ and $\rr{a}\in \bb{O}$.
However, the equality $\mu_\ast(f)\omega_\ast(\rr{a})=\wp(\mu_\ast,\omega_\ast)(f\odot \rr{a})$ on every simple tensor defines uniquely the states $\wp(\mu_\ast,\omega_\ast)$. This shows that $\wp(\mu_\alpha,\omega_\beta)$ converges to $\wp(\mu_\ast,\omega_\ast)$ in the $\ast$-weak topology, proving the continuity of $\wp$.
The other implication is even easier. Let $\{\eta_\alpha\}\subset\s{P}_{C(X)\otimes\bb{O}}$ a net converging to $\eta_\ast\in\s{P}_{C(X)\otimes\bb{O}}$. One has that $\wp^{-1}(\eta_\alpha)=(\mu_{\eta,\alpha},\omega_{\eta,\alpha})$ where $\omega_{\eta,\alpha}(\rr{a}):=\eta_\alpha(1\odot \rr{a})$ by definition.
Therefore  $\omega_{\eta,\alpha}(\rr{a})$ converges to 
$\eta_\ast(1\odot \rr{a})$ for evry $\rr{a}\in\bb{O}$. A similar argument (with the use of an approximated identity) shows that 
$\mu_{\eta,\alpha}(f)$ converges to 
$\lim_{n\to\infty}\eta_\ast(f\odot \rr{e}_n)$ for every $f\in C(X)$. It turns out that $\wp^{-1}(\eta_\alpha)$ converges exactly to $\wp^{-1}(\eta_\ast)$ in the $\ast$-weak topology of the cartesian product. The rest of the proof follows by the natural homeomorphism 
$\s{P}_{\bb{A}}\simeq \s{P}_{C(X)\otimes\bb{O}}$ induced by the isomorphism of the related $C^*$-algebras and from the homeomorphism 
in Lemma \ref{lem:homCX}.
\qed

%------%
\subsection{Proofs of the main theorems}\label{sec:th_proof}
The main purpose of this section is to provide the proof of
Theorems \ref{th:main1} and  \ref{th:main2}. Nevertheless, we will start by proving a technical result needed  to justify 
Definition \ref{def:LC}.
\begin{proposition}\label{pro inj}
Let $\bb{A}$ a $C^*$-bundle over $X$ with localizing homomorphisms $\pi_x:\bb{A}\to \bb{A}_x$. Let $\omega_x\in \s{P}_{\bb{A}_x}$  and consider the \emph{lift} map $\rr{i}(\omega_x):=\omega_x\circ\pi_x$. Then, 
$\rr{i}:\s{P}_{\bb{A}_x}\to \s{P}_{\bb{A}}$ 
is an injection.
\end{proposition}
\proof
Assume that $\rr{i}(\omega_x)$
 is not pure. Then there are two distinct elements $\widetilde\omega_1,\widetilde\omega_2\in \s{S}_{\bb{A}}$ and a $0<t<1$ such that 
\[
\rr{i}(\omega_x)(\rr{a})\;=\;t\widetilde\omega_1(\rr{a})+(1-t)\widetilde\omega_2(\rr{a})\;, \qquad \forall\; \rr{a}\in\bb{A}\;.
\]
Since $\rr{i}(\omega_x)(\rr{a})=0$ for every $\rr{a}\in \bb{I}_x$
one gets that 
\[
\widetilde\omega_1(\rr{a})\;=\;-\frac{(1-t)}{t}\;\widetilde\omega_2(\rr{a})\;,\qquad \forall \rr{a}\in \bb{I}_x\;.
\]
The last equality shows that $\widetilde\omega_1(\rr{a})=\widetilde\omega_2(\rr{a})=0$ for every positive element $\rr{a}\geqslant 0$ in the ideal $\bb{I}_x$. Since an ideal is generated by its positive elements one deduces that 
$\widetilde\omega_1(\rr{a})=\widetilde\omega_2(\rr{a})=0$ for every 
$\rr{a}\in \bb{I}_x$. Therefore one gets that  $\widetilde\omega_1$ defines a $\omega_1\in \s{S}_{\bb{A}_x}$ by means of the prescription 
$\omega_1(\pi_x(\rr{a}))=\widetilde\omega_1(\rr{a})$. The same holds for $\widetilde\omega_1$. But this leads to
\[
\omega_x(\rr{a}')\;=\;t \omega_1(\rr{a}')+(1-t) \omega_2(\rr{a}')\;, \qquad \forall\; \rr{a}'\in\bb{A}_x 
\]
and this contradicts the purity of $\omega_x$.
Then one gets that $\rr{i}(\omega_x)\in \s{P}_{\bb{A}}$.
If $\rr{i}(\omega_x)=\rr{i}(\omega_x')$ for $\omega_x,\omega_x'\in \s{S}_{\bb{A}_x}$, then $\omega_x(\pi_x(\rr{a}))=\omega_x'(\pi_x(\rr{a}))$ for every $\rr{a}\in \bb{A}$, and this implies that  $\omega_x=\omega_x'$. Therefore the map is injective.
\qed

\medskip

Let $\bb{A}\simeq C(X)\otimes\bb{O}$ be a trivial $C^*$-bundles. Then, 
one has the following bijective identifications
\begin{equation}\label{eq:mani-isom}
C(X,\s{P}_{\bb{A}})\;\simeq\; C(X,X\times \s{P}_{\bb{O}})\;\simeq\;
C(X,X) \times C(X,\s{P}_{\bb{O}})\;.
\end{equation}
The first identification is a direct consequence of the homeomorphism of  Proposition \ref{prop:prod_st}. The second identification follows from the fact that if $F\in C(X,X\times \s{P}_{\bb{O}})$, then $F(x)=(f_1(x),f_2(x))$ where $f_j(x)=({\rm pr}_j\circ F)(x)$, $j=1,2$, 
in function of the canonical projections 
${\rm pr}_1: X\times \s{P}_{\bb{O}}\to X$ and  ${\rm pr}_2: X\times \s{P}_{\bb{O}}\to \s{P}_{\bb{O}}$.
Let  
$C_\s{L}(X,\s{P}_{\bb{A}})$ be the set of  continuous maps described in Definition  \ref{def:LC}.
The following result is a consequence of Proposition \ref{pro inj}.
\begin{proposition}\label{prop_idennt_OO}
There is a bijective identification 
\begin{equation}\label{eq:bij_funct}
C_\s{L}(X,\s{P}_{\bb{A}})\;\simeq\; C(X,\s{P}_{\bb{O}})\;.
\end{equation} 
\end{proposition}
\proof
Consider the trivial (product) bundle ${\rm pr}_1: X\times \s{P}_{\bb{O}}\to X$. From the the  homeomorphism of  Proposition \ref{prop:prod_st} one can deduces that 
$$
\rr{i}(\s{P}_{\bb{A}_x})\;\simeq\;\{x\}\times \s{P}_{\bb{O}}\;=\;{\rm pr}_1^{-1}(x) 
$$ for every $x\in X$, where $\rr{i}$ is the lift map which enters in  Definition \ref{def:LC}. Therefore, the condition 
$F\in C_\s{L}(X,\s{P}_{\bb{A}})$ translates to $F(x)\in {\rm pr}_1^{-1}(x)$ for every $x\in X$, or equivalently to ${\rm pr}_1\circ F={\rm Id}_X$. In other words $F$ is a \emph{section} of the product bundle $X\times \s{P}_{\bb{O}}$, and in particular $F(x) =(x, f_2(x))$ for every $x\in X$. Therefore, the bijection \eqref{eq:bij_funct} is provided by the fact that $F\in C_\s{L}(X,\s{P}_{\bb{A}})$ if an only if $F=({\rm Id}_X, f_2)$ for a given $f_2\in C(X,\s{P}_{\bb{O}})$.\qed

\medskip

We are naw in position to provide the proof of   our first main result.
For that, let us recall that if $Y$ and $Y'$ are homeomorphic topological spaces then $[X,Y]\simeq[X,Y']$ for every topological space $X$.
\proof[Proof of Theorem \ref{th:main1}]
The homeomorphism in Proposition \ref{prop:prod_st} implies that
\[
{\rm TF}(\bb{A})\;:=\;[X, \s{P}_{\bb{A}}]\;\simeq\;[X,X\times \s{P}_{\bb{O}}]
\]
and the identification \eqref{eq:mani-isom} induces the bijection 
\[
[X,X\times \s{P}_{\bb{O}}]\;\simeq\;[X,X]
\times [X,\s{P}_{\bb{O}}]\;.
\]
Finally, one infers that 
\[
{\rm LTF}(\bb{A})\;:=\;[X,\s{P}_{\bb{A}}]_\s{L}\;\simeq\;[X,\s{P}_{\bb{O}}]
\]
in view of Proposition \ref{prop_idennt_OO}.
\qed

\medskip

For the next proof we need to recall that to spaces $Y$ and $Y'$ are 
\emph{weak homotopy equivalent} if there is a continuous map $\imath:Y\to Y'$ which induces isomorphisms of all the homotopy groups, \ie $\imath_*:\pi_k(Y)\to \pi_k(Y')$ is an isomorphism for every $k\in\N \cup\{0\}$. We will use the notation $Y \thicksim Y'$ for  
weak homotopy equivalent spaces.
In the relevant case that  $X$ is (homotopy equivalent to) a CW-complex, then $Y \thicksim Y'$ implies that $[X,Y]\simeq [X,Y']$ \cite[Theorem 2]{matumoto-minami-sugawara-84}.  

\proof[Proof of Theorem \ref{th:main2}]
In view of Theorem \ref{th:main1} and Lemma \ref{lem:comp} one obtains
\[
{\rm LTF}(\bb{A})\;\simeq\;[X,\s{P}_{\bb{K}}]\;\simeq\;[X,\n{P}(\s{H})_w]\;.
\] 
To complete the proof it is sufficient to notice that  $\n{P}(\s{H})_w$ is weak homotopy equivalent to the space $B\n{U}(1)$
 which is the classification space of complex line bundle over $X$
 \cite[Theorem 1.1]{denittis-gonzalez-gomi-25}. Therefore, one has that  
 \[
[X,\n{P}(\s{H})_w]\;\simeq\;[X,B\n{U}(1)]\;\simeq\;{\rm Pic}(X)
\]
where ${\rm Pic}(X)$ is the set of isomorphism classes of  complex line bundle over $X$ (also known as Picard group). The final isomorphism  
${\rm Pic}(X){\simeq}H^2(X,\Z)$ is a classical result in the classification theory of vector bundles.
\qed

\medskip

We will refer to the standard textbooks \cite{atiyha-67,husemoller-94,hatcher-17} for all the necessary information about the classification theory of vector bundles, classifying spaces and characteristic classes.

\proof[Proof of Corollary \ref{coro:int}]
The crucial observation is the \emph{stable range theorem} \cite[Theorem 1.2, Chapter 9]{husemoller-94} which implies that  
the classification of vector bundles of any rank reduces to the classification of line bundles if the dimension of $X$ is bounded by $3$. This means that ${\rm Vec}_\C^m(X)\simeq {\rm Pic}(X)$ for every rank $m$, and under the canonical inclusion ${\rm Vec}_\C^m(X)\hookrightarrow {\rm Vec}_\C^{m+1}(X)$ given by the sum   
of a trivial line bundle one gets ${\rm Vec}_\C^\bullet(X) \simeq {\rm Pic}(X)$ where ${\rm Vec}_\C^\bullet(X)$ is the set of equivalence classes of complex vector bundles under \emph{stable} isomorphisms. 
However, the latter is classified exactly by the reduced $K$-theory of $X$, \ie ${\rm Vec}_\C^\bullet(X)\simeq \widetilde{K}^0(X)$.
\qed

%------------------------------------------------------------------------%
\section{Application to relevant systems}
\label{sec:appl_syst}

\subsection{Crossed product $C^*$-algebras and the Green's theorem}\label{sec_green}
Crossed product $C^*$-algebras play a central rol in many areas of mathematical physics. For the general theory of these mathematical objects we will refer to the modern monograph \cite{williams-07} and references therein. The connection of crossed product $C^*$-algebras
with quantum systems and condensed matter systems has been largely studied by various authors, see for instance to \cite{Bellissard-93,bellissard-elst-schulz-baldes-94,lenz-99,georgescu-iftimovici-02,georgescu-iftimovici-06,mesland-prodan-22} just to mention a few of them. 

\medskip

Let $\n{G}$ be a locally compact group and $\n{H}\subset \n{G}$ a closed subgroup. Consider the quotient space $\Sigma:=\n{G}/\n{H}$
which is always locally compact and Hausdorff  in the quotient topology.  The group $\n{G}$ acts on $\Sigma$ by
left-translations. More precisely if $\sigma_{g'}:=[g']=\{g'h\;|\; h\in \n{H}\}$ is a point of $\Sigma$ and $g\in\n{G}$, then $g\cdot\sigma_{g'}:=\sigma_{gg'}=[gg`]$. Consider the commutative $C^*$-algebra $C_0(\Sigma)$
on which $\n{G}$ actas according to $\tau_g(f)(\sigma)=f(g^{-1}\cdot\sigma)$ for every $f\in C_0(\Sigma)$, $\sigma\in\Sigma$ and $g\in\n{G}$.
The space of compactly supported functions
$C_c(\n{G},C_0(\Sigma))$  can be endowed with the convolution product
\[
(F\ast G)(g)\;:=\;\int_{\n{G}}\dd\mu(h)\; F(h)\tau_h(G)(h^{-1}g)
\]
and the involution
\[
F^*(g)\;:=\;\Delta(g^{-1})\tau_g\left(\overline{F(g^{-1})}\right)
\]
for every $F,G\in C_c(\n{G},C_0(\Sigma))$, where $\mu$ is any Haar measure with Haar modulus $\Delta$. The $\ast$-algebra  $C_c(\n{G},C_0(\Sigma))$ can be closed with respect to the \emph{universal norm} $\|\cdot\|$ to provide a $C^*$-algebra $C_0(\Sigma)\rtimes \n{G}$ called the \emph{crossed product} of $C_0(\Sigma)$ by $\n{G}$
\cite[Lemma 2.27]{williams-07}.

\medskip

It is worth noting that $\Sigma$ is a group only when $\n{H}$ is normal. If it is not the case $\Sigma$ need not possess a measure which is invariant for the left action of $\n{G}$. However, in general 
$\Sigma$ has a \emph{quasi-invariant} Borel measure $\mu_\Sigma$ such that for every $f\in C_c(\n{G})$ one has 
\[
\int_{\n{G}}\dd\mu(g)\rho(g) f(g)\;=\;\int_\Sigma\dd\mu_\Sigma(\sigma_g)
\left(\int_\n{H}\dd\mu_{\n{H}}(h)\; f(gh)\right)\]
where $\mu_{\n{H}}$ is a Haar measure on $\n{H}$ and $\rho:\n{G}\to(0,\infty)$ is a suitable function (see \cite[Section 4.5]{williams-07} for more details).
Let $\rr{h}_\Sigma:=L^2(\Sigma, \mu_\Sigma)$. The relevant fact for this work is provided by the \emph{Green's theorem} \cite[Theorem 4.30]{williams-07} which establishes the isomorphism
\begin{equation}\label{eq:cross1}
C_0(\Sigma)\rtimes \n{G}\;\simeq\; C^*(\n{H})\otimes \bb{K}(\rr{h}_\Sigma)\;,
\end{equation}
where $C^*(\n{H})$ denotes the \emph{group $C^*$-algebra} of $\n{H}$
\cite[Example 2.33]{williams-07}.

\medskip

In the case $\n{H}$ is an abelian group then  $C^*(\n{H})$ is an abelian $C^*$-algebra and
  the Gelfand isomorphism provides
$C^*(\n{H})\simeq C_0(\widehat{\n{H}})$ where $\widehat{\n{H}}$
is the \emph{Pontryagin dual group} of $\n{H}$. In this particular case \eqref{eq:cross1} reduces to 
\begin{equation}\label{eq:cross2}
C_0(\Sigma)\rtimes \n{G}\;\simeq\; C_0\big(\widehat{\n{H}}\big)\otimes \bb{K}(\rr{h}_\Sigma)\;,
\end{equation}
showing that the crossed product is in fact isomorphic to a trivial $C^*$-bundle over $\widehat{\n{H}}$ with fiber $\bb{K}(\rr{h}_\Sigma)$.
Therefore, under the assumption above, if moreover  $\n{H}$ is abelian
 and discrete (implying  that $C^*(\n{H})$ is unital) and $\rr{h}_\Sigma$ is separable, then
Theorem \ref{th:main2}
provides that
\[
{\rm LTF}(C_0(\Sigma)\rtimes \n{G})\;\simeq\;H^2\big(\widehat{\n{H}},\Z\big)\;.
\]

\subsection{Bloch $C^*$-algebra}
We will discuss now a concrete realization of the framework presented in the previous section and inspired to works like \cite{Bellissard-93,georgescu-iftimovici-02}. 

\medskip

Let $\n{G}\equiv\R^d$ and $\n{H}\equiv\Gamma\simeq\Z^d$ 
a full lattice in $\R^d$. The quotient space $\Sigma=\R^d/\Gamma$, indeed a compact group,  is called the \emph{unit cell}. 
It possesses a unique normalized Haar measure $\mu_\Sigma$ and the associated Hilbert space $\rr{h}_\Sigma:=L^2(\Sigma, \mu_\Sigma)$ is separable.
We will refer to the  abstract $C^*$-algebra
\[
\bb{A}_{\rm Bloch}\;:=\;C(\Sigma)\rtimes \R^d
\] 
as the Bloch $C^*$-algebra. The reason of this terminology is the due to the image of  $\bb{A}_{\rm Bloch}$ in the $C^*$-algebra of bounded operator $\bb{B}(L^2(\R^d))$ via the left-regular representation $\pi$. Given a $F\in C_c(\n{R}^d,C(\Sigma))\subset\bb{A}_{\rm Bloch}$, its $\pi$-representation acts on $\psi\in L^2(\R^d)$ according to 
\[
(\pi(F)\psi)(x)\;:=\;\int_{\R^d}\dd y\;\tau_{x}(F)(y-x)\psi(y)\;.
\]
Let $\gamma\in\Gamma$ a point of the lattice and $(U_\gamma\psi)(x):=\psi(x-\gamma)$ the unitary operator in $L^2(\R^d)$ that implements the translation by $\gamma$. A direct check shows that 
$U_\gamma \pi(F)U_\gamma^*=\pi(F)$ for every $\gamma\in\Gamma$.
This means that $\pi(\bb{A}_{\rm Bloch})$ is a sub-algebra of 
$\bb{B}(L^2(\R^d))$ made of $\Gamma$-invariant operators. In particular, $\pi(\bb{A}_{\rm Bloch})$ contains the resolvents of the Bloch-type operators $H_\Gamma:=-\Delta+V_\Gamma$, for $\Gamma$-periodic potentials $V_\Gamma\in L^1(\R^d)$.

\medskip

In this situation the Green’s theorem (in the version \eqref{eq:cross2}) provides
\[
\bb{A}_{\rm Bloch}\;\simeq\; C(\n{B})\otimes \bb{K}(\rr{h}_\Sigma)
\]
where $\n{B}:=\widehat{\Gamma}\simeq\n{T}^d$ is called the \emph{Brillouin torus}. Therefore, Theorem \ref{th:main2}
provides
\[
{\rm LTF}(\bb{A}_{\rm Bloch})\;\simeq\;H^2(\n{T}^d,\Z)\;=\;\Z^{\oplus{{d}\choose{2}}}
\]
which reproduces the classification of topological phases for \emph{type A} topological insulators \cite{altland-zirnbauer-97,kitaev-09,ryu-schnyder-furusaki-ludwig-10}.

\subsection{$C^*$-dynamical systems}
The next example is very frequent in mathematical physics and in the theory of dynamical systems.

\medskip

Let $\bb{A}$ be a $C^*$-algebra, $\bb{M}(\bb{A})$ its multiplier algebra and $\bb{U}(\bb{A})\subset\bb{M}(\bb{A})$ the group of the unitary elements in $\bb{M}(\bb{A})$.  Let $\n{G}$ be a locally compact group and $\rr{u}:\n{G}\to \bb{U}(\bb{A})$ a strictly continuous homomorphism. Consider the $\n{G}$-action $\tau:\n{G}\to {\rm Aut} (\bb{A})$ defined by $\tau_g(\rr{a}):=\rr{u}_g\rr{a}\rr{u}_g^*$
for every $\rr{a}\in\bb{A}$ and $g\in\n{G}$. The triple $(\bb{A},\n{G},\tau)$ is a \emph{$C^*$-dynamical system}.

\medskip

Starting with $C_c(\n{G},\bb{A})$ one can form the crossed product $C^*$-algebra $\bb{A}\rtimes \n{G}$ following the prescription described in Section \ref{sec_green}. From \cite[Lemma 2.73]{williams-07} one gets that 
\[
\bb{A}\rtimes \n{G}\;\simeq\; C^*(\n{G})\otimes_{\rm max}\bb{A}
\]
where $\otimes_{\rm max}$ denotes the \emph{maximal} tensor product.
In the case that $\n{G}$ is an abelian group then $C^*(\n{G})$ is an abelian $C^*$-algebra (hence nuclear) isomorphic  to $C(\widehat{\n{G}})$ 
via the Gelfand isomorphism where $\widehat{\n{G}}$ denotes the 
{Pontryagin dual group} of $\n{G}$. Therefore in this case one has that
\[
\bb{A}\rtimes \n{G}\;\simeq\; C(\widehat{\n{G}})\otimes\bb{A}
\]
independently of the definition of the tensor product, and
Theorem \ref{th:main1}
provides 
\[
{\rm LTF}(\bb{A}\rtimes \n{G})\;\simeq\;\big[\widehat{\n{G}},\s{P}_{\bb{A}}\big]\;.
\]

\subsection{The non-commutative torus}
The \emph{non-commutative torus} plays a relevant role in the many areas of the non-commutative geometry and mathematical physics. For its connection with the modern theory of topological insulator we will refer to \cite{prodan-schulz-baldes-book} and references therein. A complete monograph on its mathematical properties is \cite{boca-01}.

\medskip

Consider the  {$C^*$-dynamical system} $(C(\n{S}^1),\n{Z},\tau)$ where $\n{S}^1$ is parametrized as the group  $\{z\in\C\;|\;|z|=1\}$
and the $\Z$-action is implemented on $f\in C(\n{S}^1)$ by $(\tau_n f)(z):=f(\rho_\theta^n z)$ where $\rho_\theta:=\expo{-\ii 2\pi \theta}$
and $\theta\in\R$. The associated crossed product $\bb{A}_\theta:= C(\n{S}^1)\rtimes \n{Z}$ is called  \emph{rotation algebra} or also  {non-commutative torus}
\cite[Examples 2.13 \& 2.55]{williams-07}. It is also true that 
$\bb{A}_\theta$ can be realized as the \emph{universal} $C^*$-algebra generated by two elements $\rr{u}$ and $\rr{v}$ such that 
 $\rr{u}^{-1}= \rr{u}^*$,  $\rr{v}^{-1}= \rr{v}^*$ and $\rr{u}\rr{v}= \rho_\theta\rr{v}\rr{u}$
\cite[Proposition 2.56]{williams-07}.

\medskip

The properties of $\bb{A}_\theta$ strongly depend of the nature of $\theta$ being \emph{rational} or \emph{irrational}. For $\theta\in\n{Q}$ a rational number one can prove that 
\[
\bb{A}_\theta\;\simeq\; C(\n{T}^2)\otimes\bb{K}
\]
where $\n{T}^2$ is a two-dimensional torus and $\bb{K}$ is the $C^*$-algebra of compact operators on some separable Hilbert space 
\cite[Remark 2.60 \& Example 8.46]{williams-07}. 
Therefore for any rational $\theta\in\n{Q}$ 
 Theorem \ref{th:main2}
provides
\[
{\rm LTF}(\bb{A}_\theta)\;\simeq\;H^2(\n{T}^2,\Z)\;=\;\Z\;.
\]
Since the algebra $\bb{A}_\theta$ is used to model bidimensional discrete magnetic systems, the result above reproduces the classification of distinct \emph{Hall phases} \cite{bellissard-elst-schulz-baldes-94}.

\end{document}